\documentclass{article}

\usepackage[preprint]{neurips_2024}

\usepackage[utf8]{inputenc} % allow utf-8 input
\usepackage[T1]{fontenc}    % use 8-bit T1 fonts
\usepackage{hyperref}       % hyperlinks
\usepackage{url}            % simple URL typesetting
\usepackage{booktabs}       % professional-quality tables
\usepackage{amsfonts}       % blackboard math symbols
\usepackage{nicefrac}       % compact symbols for 1/2, etc.
\usepackage{microtype}      % microtypography
\usepackage{xcolor}
\usepackage{graphicx}
\usepackage{ifthen} 
\usepackage{wrapfig}
\usepackage{natbib}
\usepackage{amsmath}
\usepackage{amsthm}
\newtheorem{definition}{Definition}

\DeclareMathOperator*{\argmax}{arg\,max}

\usepackage{listings}

\definecolor{codebg}{rgb}{0.96,0.96,0.96}
\definecolor{codeframe}{rgb}{0.85,0.85,0.85}

\lstdefinestyle{customc}{
  backgroundcolor=\color{codebg},       % background color
  breaklines=true,                      % enable line breaking
  frame=single,                         % frame style
  rulecolor=\color{codeframe},          % frame color
  xleftmargin=\parindent,               % left margin
  language=Bash,                        % language (bash in this case)
  showstringspaces=false,               % do not show spaces in strings as special characters
  basicstyle=\footnotesize\ttfamily,    % basic style
  keywordstyle=\text{},   % keyword style
  commentstyle=\itshape\color{purple},  % comment style
  identifierstyle=\color{black},        % identifier style
  numbers=none,                         % line numbers style
  mathescape=true,                      % allows escaping to math mode in $
  escapechar=@                          % escape character for LaTeX within the listing
}

\newboolean{showcomments}
\setboolean{showcomments}{true}  % Set to true to show comments, false to hide them

\newcommand\todo[1]{\ifthenelse{\boolean{showcomments}}{\textcolor{red}{TODO: #1}}{}}
\newcommand\atrisha[1]{\ifthenelse{\boolean{showcomments}}{\textcolor{blue}{Atrisha says: #1}}{}}
\newcommand\carter[1]{\ifthenelse{\boolean{showcomments}}{\textcolor{purple}{Carter says: #1}}{}}
\newcommand\andrei[1]{\ifthenelse{\boolean{showcomments}}{\textcolor{orange}{Andrei says: #1}}{}}
\newcommand\gillian[1]{\ifthenelse{\boolean{showcomments}}{\textcolor{teal}{Gillian says: #1}}{}}
\newcommand\rakshit[1]{\ifthenelse{\boolean{showcomments}}{\textcolor{magenta}{Rakshit says: #1}}{}}

\title{Normative Modules: A Generative Agent Architecture for Learning Norms that Supports Multi-Agent Cooperation}

\author{%
  Atrisha Sarkar \\
  Schwartz Reisman Institute for Tech. and Society\\
  Vector Institute for Artificial Intelligence\\
  University of Toronto\\
  \texttt{atrisha.sarkar@utoronto.ca}
   \And
   Andrei Ioan Muresanu \\
   Vector Institute for Artificial Intelligence\\
  University of Waterloo\\
  \texttt{andrei.muresanu@uwaterloo.ca} \\
  \And
   Carter Blair \\
   Vector Institute for Artificial Intelligence\\
  University of Waterloo\\
  \texttt{cblair@uwaterloo.ca} \\
  \And
   Aaryam Sharma \\
  University of Waterloo\\
  \texttt{a584sharma@uwaterloo.ca} \\
  \And
   Rakshit S Trivedi \\
  Massachusetts Institute of Technology\\
  \texttt{rstrivedi@csail.mit.edu} \\
  \And
   Gillian K Hadfield \\
   Schwartz Reisman Institute for Tech. and Society\\
   Vector Institute for Artificial Intelligence\\
  University of Toronto\\
  \texttt{g.hadfield@utoronto.ca} \\
}

\begin{document}

\maketitle

\begin{abstract}
Generative agents, which implement behaviors using a large language model (LLM) to interpret and evaluate an environment, has demonstrated the capacity to solve complex tasks across many social and technological domains. However, when these agents interact with other agents and humans in presence of social structures such as existing norms, fostering cooperation between them emerges as a fundamental challenge.
%- a problem that presents challenges even in traditional multiagent reinforcement learning (MARL) environments. 
In this paper, we develop the framework of a \textbf{Normative Module}: an architecture for generative agents designed to enhance cooperation by enabling agents to recognize and adapt to the normative infrastructure of a given environment, in the form of institutions that define acceptable behaviors within a group of agents.
%we develop the framework of a \textit{normative module}: an architecture for generative agents that imparts the capacity for cooperation in institutional environments by developing agents’ capacity to recognize normative infrastructure, specifically institutions that encode the behaviors that a group of agents treats as sanctionable. 
%We call such institutions “classification institutions” (in reference to the human practice of designating a common-knowledge location where a group’s operative rules can be discovered.) 
We focus on the \textit{equilibrium selection} aspect of the cooperation problem and inform our agent design based on the existence of \textit{classification institutions} that implement correlated equilibrium to provide effective resolution of the equilibrium selection problem.
%ground the problem in a game-theoretic framework in which an classification institution implements a correlated equilibrium in a sanctioning game. 
Specifically, the normative module enables agents to learn through peer interactions which of multiple candidate institutions in the environment, does a group treat as authoritative. By enabling normative competence in this sense, agents gain ability to coordinate their sanctioning behaviour; coordinated sanctioning behaviour in turn shapes primary behaviour within a social environment, leading to higher average welfare 
%(Koster et al 2022). 
We design a new environment that supports institutions and evaluate the proposed framework based on two key criteria derived from agent interactions with peers and institutions: (i) the agent's ability to disregard non-authoritative institutions and (ii) the agent's ability to identify authoritative institutions among several options. Crucially, we show that these capabilities allow the agent to achieve more stable cooperative outcomes compared to baseline agents without the normative module, paving the way for future research in a new avenue of designing environments and agents that account for normative infrastructure.

\end{abstract}

% \input{sections/1_intro}
% \input{sections/2_related_work}
% \input{sections/3_background}
% \input{sections/4_architecture}
% %\input{sections/5_environment}
% \input{sections/6_experiments}
% \input{sections/7_discussion}
% \input{sections/8_appendix}

\section{Introduction}

The unprecedented reasoning, planning, and communication ability of the current generation of LLM-based foundation models has led to a new class of autonomous generative agents that can perform complex tasks across domains in an open-world multi-agent environment. These include planning the high-level behaviour of autonomous vehicles \citep{yang2023survey}, human-robot interaction \citep{mahadevan2024generative}, providing mental health support \citep{ma2023understanding}, engaging in market activity as economic agents \citep{horton2023large}, solving collective action problems \citep{park2023generative}, and simulating political behaviour \citep{argyle2023out}. From a computational perspective, the primary goal for these applications is building agent architectures that can both simulate human and social behaviour and seamlessly interact with humans.
However, such open-world interaction raises a key challenge -- human behaviour, individually and collectively, is not determined solely through the kind of goal-orientated task solving that has guided the architecture of the current state-of-the-art autonomous agents. What sets human societies apart from other mammals is the human capacity to engage in extensively cooperative behaviour \cite{boyd2009culture}. Human tasks, like driving or work or political behavior, are embedded in dense normative social orders \cite{hadfield2014microfoundations}: stable states in which defined groups (a workplace, a neighborhood, a country, an ethnic group) have achieved coordinated third-party enforcement of and hence general compliance with a public and shared set of norms. Indeed, for many groups, these norms define what it means to be a member of the group: we facilitate merging on the highway, we bluntly tell a co-worker that they have executed a task poorly, we don't cooperate with legislators from the opposing party. 

\citep{hadfield2012law} propose a parsimonious rational agent model of normative social order, consisting of a \textit{classification institution} that provides common knowledge binary classification of all behaviors as either "punishable" or "not punishable" (possibly through complex application of general principles to particular cases) and an \textit{enforcement mechanism} that incentivizes agents to prefer "not punishable" actions. A stable normative social order is achieved when most agents are mostly in compliance and avoiding punishment. \citep{hadfield2012law} focus in particular on the case, which describes most of human history and much of modern life as well, in which punishment is primarily delivered by ordinary agents (rather than specialized enforcers \cite{hadfield2013law}.) Agents must therefore be incentivized and coordinated to engage in costly third-party punishment (which could be relatively mild, such as criticism, or more harsh, such as exclusion from the group) and to condition such punishment actions on a shared classification institution. Although shared classification could be entirely emergent and informal--there is no entity that tells members of the group that they should honk at the car that is failing to facilitate merging on the highway but everyone in the group could reliably say that this is the norm \cite{Bicchieri2005TheGO} -- groups that converge on a single authoritative (more formal) classification institution--such as a chief, a group of elders, or a court--can enjoy the benefits of maintaining cooperation even in the face of changing environments and populations \cite{hadfield2017rules}. For autonomous generative agents to effectively operate in open-world human-like environments, they must possess \textit{normative competence}: the capacity to recognize and reason about the norms and institutions (such as classification institutions) that condition a population's normative behavior and the capacity and incentive to engage in (punish) and respond to (comply) sanctioning, conditioned on shared classifications and the behavior of others. The goal of building agents for general-purpose human social behaviour thus raises the following fundamental question: 

\begin{center}
\textit{ How to design autonomous agent architectures with the capacity for reasoning about normative institutions and behaviors within a social environment?}\par
\end{center}

While there have been promising attempts towards addressing this question through the lens of reinforcement learning agents in multi-agent environments \citep{vinitsky2023learning, koster2022spurious, leibo2021scalable}, this question has been largely under-explored in the context of generative agents. To bridge this conspicuous gap and address this question for generative agents, we deviate from previous literature on a key dimension:  we shift the design challenge from training agents to punish and comply with designer-established norms and rewards for punishment to the more generalizable challenge of training a new-comer to learn the normative infrastructure and normative state of an established group of agents. As a starting point, we formalize the problem of resolving cooperation through normative infrastructure. Next, we imbue generative agents with a novel \textbf{Normative Module}, that enables the agent to reason about the classification institutions in the environment. Specifically, the normative module performs two functions: (i) learning about the authoritative classification institutions in a community from community interactions; and (ii) selecting actions from the candidate set based on the cost associated with an internal representation of the sanctions from an authoritative institution. 
Finally, we design a new environment, called \textit{Normative Orchards}, where a group of background agents and a foreground agent play a fruit harvesting game. The environment consists of variations of classification institutions that approve or disapprove of appropriate harvesting behaviour. This environment poses an equilibrium selection problem to the generative agents, and agents need to reason about both classification institutions and behaviours of other agents towards achieving stable cooperative outcomes.

In our experiments, we investigate these implications by asking the following questions:

\textit{(i) When confronted with a single non-authoritative classification institution, are generative agents with a normative module adept at recognizing the unreliability of that institution and thereby aligning with the behavior of other agents in the community?}

\textit{(ii) Given a set of multiple classification institutions, do generative agents with a normative module recognize the authoritative institution correctly and align their behavior with that institution so as to engage in cooperation?}

\textit{(iii) Do generative agents imbued with normative module demonstrate behavioral distinctions towards achieving a cooperative outcome, as compared to baseline models without the module?}
Our empirical results demonstrate that generative agents that possess what we call a  \textit{normative module} can, in comparison to a generative agent that lacks such a module, more rapidly identify which of multiple classification institutions in an environment is authoritative for a group of (hard-programmed) background agents and better align behavior with the group's established behaviors. If the only classification institution in the environment is not treated as authoritative by the group--meaning the background agents are not following the norm announced by the institution--the normative module agent is also better able to determine this and ignore the institution.

\section{Related Work}

There is a large body of work addressing the challenge of cooperation in mixed-motive multi-agent learning (MAL). Solutions include centralized training \citep{baker2020emergent, foerster2018counterfactual, iqbal2019coordinated, lowe2017multi}, formal contracts \citep{christoffersen2022get, yocum2023mitigating}, opponent shaping \citep{lu2022model,khan2023scaling}, gifting \citep{lupu2020gifting, wang2021emergent}, and the optimization of social influence \citep{jaques2019social, meta2022human}. Centralized training approaches have been particularly successful in bringing about cooperation, but the assumption that all agents can be trained in a centralized manner may not hold in realistic settings. For an in-depth review of cooperation in multi-agent learning, see \citep{du2023review}.

Social norms have long been recognized as useful for promoting and stabilizing cooperation in multi-agent systems \cite{gintis2010social, hadfield2012law, vanderschraaf1995convention} and have also been used to promote cooperation in mixed-motive settings multi-agent learning \cite{vinitsky2023learning, koster2022spurious}. In the work done by \citet{vinitsky2023learning}, agents learn social norms through public sanctioning, where behaviours are classified based on the observed approval and disapproval of other agents. This enables agents to develop personal representations of acceptable behaviour, resulting in a "bandwagon" effect that stabilizes norms. This approach relies on the sanctions of each agent in the public environment so that each agent may learn a classifier of acceptable behaviour. However, in reality, sanction observations may not be public and thus may be sparse, making learning such a classifier difficult. 

Institutions also play an important role in normative behaviour \cite{hadfield2012law, aoki1996towards}. However, previous works using social norms to promote cooperation in mixed-motive multi-agent learning have largely ignored this aspect. \citet{aoki1996towards} emphasizes that institutions and norms are interdependent, forming a coherent system that evolves and adapts. \citet{hadfield2012law} highlight the role of decentralized enforcement and the need for coordination and a common logic to maintain legal order. They argue that institutions can provide a common logic for classifying behaviours as wrongful or not, which helps coordinate the normative actions of individuals. These insights suggest that incorporating institutional mechanisms could enhance the stability and effectiveness of normative behaviours in multi-agent systems. Integrating such mechanisms could address the challenge of sparse sanction observations and enhance the overall robustness and convergence time of cooperation-promoting strategies.

More recently, some work has been done using generative (LLM-based) agents in mixed-motive settings. Work by \citet{akata2023playing} explored using LLMs to play repeated games, finding that LLM agents were generally capable in tasks like iterated Prisoner's Dilemma but weaker in coordination games like Battle of the Sexes. Further, \citet{fan2024can} analyzed LLM's reasoning ability from a game theoretic perspective, finding that they struggle in refining their beliefs over time based on observed patterns.  
\citet{yocum2023mitigating} specifically looked at generative agents in social dilemma settings and found that generative agents could successfully negotiate contracts that promote cooperation.

\section{Resolving Cooperation Challenges through Normative Infrastructures}

%Designing agents that can resolve social dilemmas is a key question in multiagent systems. In such a setting, a conflict exists between individual incentives, encoded in each player's utility function, and the collective group benefit, encoded through an aggregation of the utilities of all players in the environment. Norm-based resolution through third-party sanction has become an increasingly popular approach for resolving social dilemmas in multiagent environments. In such an approach, agents' actions are augmented with a sanction action that the agent can apply to other agents at a cost to the other agent. In this section, we provide a game-theoretic analysis of these methods, which helps ground this approach in a theoretical framework and help us inform the design of the normative module architecture in the subsequent sections. \par
Most efforts to resolve cooperation challenges in MAL focus on direct solutions to the cooperation challenge: algorithms and agent architectures that encourage agents to learn policies that lead to more cooperative actions. But this approach misses a key insight from human cooperation. Human cooperation is highly dependent on structured third-party punishment. %A prisoner's decision not to rat out their co-conspirator is more often explained by concerns about what will happen if they violate the code of their gang than by expectations about loss of cooperation from this particular co-conspirator the next time they are arrested together. 
Explaining why humans engage in third-party punishment (which is costly to the punisher and, unlike second-order retaliation, brings no direct benefit to the punisher) is a fundamental puzzle in the literature on human origins and cultural evolution \cite{boyd1992punishment,boyd2009culture}.  But whatever the explanation for the emergence of sustained third-party punishment schemes in human groups, it is clear that coordinated punishment greatly expands, and in many cases simplifies, the solution space for cooperation challenges. Indeed, third-party punishment \textit{transforms} an underlying cooperation challenge into a coordination challenge: how can agents coordinate their punishment efforts such as to deter uncooperative behavior? \citep{hadfield2012law} theorize that this is the function of distinctively legal institutions: providing an authoritative, common knowledge classifications of behaviors to coordinate punishment on a shared set of norms.  In this section, we formalize this framework to guide the design of the normative module for generative agents.
We develop the underlying formalism over a general-sum extensive form game (EFG) $\Gamma=(N,X,T,A,\Sigma,U)$ \footnote{The reader can refer \citep{kovavrik2022rethinking} for translation between EFG and Partially Observble Stochastic Game (POSG), a structure more common in multiagent reinforcement learning environments}.  $N$ represents the finite set of players in the environment with players indexed by $i \in N$. $X$ represents the set of states that form the nodes of the game tree and includes the empty set $\emptyset$ and terminal states $Z \subset X$.  $A_{i,x}$ is the set of actions available to each player in state $x \in X$ and $A_{x} = \{A_{i,X}:i\in N\}$ is the joint set of actions. $T: X \times A \rightarrow X$ is the environment transition function that maps agents' actions to the next state.  Let $\mathcal{H}$ represent the finite set of histories in the game with each element $h_{x} \in \mathcal{H}$ representing the sequence of actions of all players from $\emptyset$ to the node $x$. Then a strategy for a player $i$ (in behavior form \citep{kuhn1953extensive}) is denoted by $\sigma_{i}:\mathcal{H} \rightarrow A_{i,x}$ which maps a history terminating at state $x$ to an action selected by the player $i$, and $\Sigma_{i}$ is the the space of all strategies that assigns an action to all histories. $u_{i}: \{h_{x\in Z}\} \rightarrow [0,1]$ is utility for the player $i$ defined at the set of terminal nodes $x \in Z$ and $U$ is the vector of utilities for all players. The cooperation dilemma in such an EFG is as follows. \par

%Although the problems of cooperation and coordination are often used collectively and interchangeably within the cooperative multi-agent literature \citep{du2023review}, we make the distinction that the problem of the social dilemma of cooperation highlighted earlier comes in two different flavours. The first presents a larger problem of the cooperation dilemma, in which there is an incentive for an agent to be non-cooperative when other agents are cooperative; therefore, any equilibrium is not the welfare-maximising one. This type of dilemma often reflects a payoff akin to a prisoner's dilemma game. The second is the coordination dilemma, in which there are multiple equilibria, some of which are the welfare-maximising solution, and agents need to converge to the welfare-maximising one. Our analysis focuses on the cooperation dilemma, since that presents a greater challenge in an environment of rational utility-maximising agents.
\begin{definition}

A cooperation dilemma for player $i$ is a game in which the following condition holds true: $\exists \sigma_{i}: u_{i}(\sigma_{i},\sigma^{\text{SW}}_{i}) > u_{i}(\sigma^{\text{SW}}_{i},\sigma^{\text{SW}}_{i})$ where $\sigma^{\text{SW}} = \argmax\limits_{\sigma=(\sigma_{i},\sigma_{-i})} \sum\limits_{i} u_{i}(\sigma_{i},\sigma_{-i})$.
\label{defi:coope_problem}
\end{definition}
\iffalse
\begin{definition}
A coordination dilemma for player $i$ is a game in which the following condition holds true $\exists -i: \sigma^{\text{maxNE}_{i}} \neq \sigma^{\text{maxNE}_{-i}}$, where  $\sigma^{\text{maxNE}_{i}} = \argmax\limits_{\sigma \in \Sigma^{\text{NE}}} u_{i}(\sigma)$ the utility maximizing Nash equilibrium strategy profile for $i$ and $\Sigma^{\text{NE}}$ is the set of Nash equilibria.
\end{definition}
\fi

According to the Folk Theorem \citep{Fudenberg1986}, if a game is repeated for an indefinite period and players are sufficiently patient, any cooperative outcome can be achieved through trigger strategies according to which players punish defectors. The challenge generated by the Folk Theorem is one of equilibrium selection, namely, coordinating all players on a specific trigger strategy (when and how much to punish defectors) from a large set of possible strategies, all of which achieve cooperation. Human societies solve this problem through normative infrastructure: establishing norms and institutions that coordinate punishment strategies \citep{Bicchieri2005TheGO,boyd1992punishment,hadfield2012law}. Drawing from this human model , current methods in multi-agent learning augment the agent's action space in the original game ($\Gamma$) with \emph{sanction} actions that agents can impose on other agents \citep{perolat2017multi,vinitsky2023learning}. These sanctions result in a cost to the targeted agent; they are also costly to the sanctioning agent and hence agents must be incentivized (either by a private benefit such as the removal of a competitor or by a reward generated from the environment for specific sanctioning actions) to engage in punishment. Provided agents are incentivized and coordinated to sanction specific non-cooperative actions, the sanction-based mechanism shapes collective behaviour towards a welfare-improving cooperative outcome. \par
We associate the game $\Gamma$ representing the cooperation dilemma with a separate \emph{sanction game} $\Gamma_{S}$. In each history $h_{x}$ of the original game $\Gamma$, a player can execute a strategy in the associated game $\Gamma_{S}$. The constructs of the sanction game include the same set of players $N$ and the actions are a set of \emph{classification functions} that each player uses to label strategy profiles in $\Gamma$ as cooperative or not.\footnote{\citep{hadfield2012law}refer to these functions as an \textit{idiosyncratic logic}.} If a player classifies an action by player $-i$ in history $h_{x}$ as non-cooperative, the player can impose a cost (set by the designer and exogenous to $\Gamma$) by sanctioning player $-i$. Therefore, the strategy profile in $\Gamma_{S}$ is the set of classification functions used by all agents, and the utilities are the sum of the costs the agents impose on each other based on the history and their classification function.
When the outcome of the utilities (cost) in $\Gamma_{S}$ is added to the original game $\Gamma$, the strategies of the sanction game $\Gamma_{S}$ generates a separable transform \citep{tewolde2021game} of the original game. The transform $H_{i}$ acts on $\Gamma$ such that $\forall \sigma \in \Sigma$, $H_{i}(U_{i}): u_{i}'(\sigma_{i},\sigma_{-i}) = u_{i}(\sigma_{i},\sigma_{-i}) - v_{i}(C_{i}^{*}(\sigma), C_{-i}^{*}(\sigma))$, where $v_{i}$ is the cost incurred by player $i$ in the sanction game $\Gamma_{S}$ and $C_{i}^{*}$ is the equilibrium sanction function in $\Gamma_{S}$ that classifies a strategy profile $\sigma$ sanctionable and $u'_{i}$ is the resulting transformed payoff for player $i$ for strategy profile $(\sigma_{i},\sigma_{-i})$ in the original game $\Gamma$.
A transform of the utilities in the original game through the sanction game is dilemma resolving if and only if the incentive to deviate to a non-cooperative strategy (as established in Defn. \ref{defi:coope_problem}) is offset by the cost imposed by others in the equilibrium of the sanction game.
\begin{definition}
    A sanction based transform is cooperation dilemma resolving iff $\exists \sigma_{i}: u_{i}(\sigma_{i},\sigma^{\text{SW}}_{i}) > u_{i}(\sigma^{\text{SW}}_{i},\sigma^{\text{SW}}_{i})$ and $ u'_{i}(\sigma_{i},\sigma^{\text{SW}}_{i}) < u'_{i}(\sigma^{\text{SW}}_{i},\sigma^{\text{SW}}_{i})$
\end{definition}
When the sanction game is played infinitely often, for example, through the affordances of the game mechanics, the space of all equilibrium costs from the sanction game can be characterized using the set of Nash equilibrium payoff using folk theorem highlighting the presence of infinitely many equilibrium possibilities. Therefore, a norm-based solution to resolving the cooperation problem for agents essentially boils down to selecting a coordinated strategy in the sanction game.

\subsection{Classification Institutions}
\label{sec:class_inst}
The first problem highlighted here is the need for coordination for a sanctioning strategy $(C_{-i}(\sigma_{i}),C_{-i}(\sigma_{-i}))$ that resolves a cooperation problem. We propose that a natural way to solve this equilibrium selection problem is through a classification institution that implements a \emph{correlated equilibrium} of the sanction game \citep{aumann1987correlated}. This translates to coordination of sanction classification by institutions, in the form of a declaration of what actions should be punished. Following that, from an agent's point of view, they cannot gain (in $\Gamma_{S}$) by deviating to another classification function (including a personal classification function) contrary to the institution's classification. Formally, let, $\mathcal{A_{\mathcal{I}}}:C_{i}^{N} \rightarrow [0,1]$ be a distribution of classification strategies in the sanction game that an institution $\mathcal{I}$ uses, then the distribution $\mathcal{A_{\mathcal{I}}}^{\text{CE}}$ implements a correlated equilibrium under the following condition:
\begin{equation}
    \mathbf{E}_{C^{*}_{-i} \sim \mathcal{A}_{I}^{\text{CE}}}[v_{i}(C^{*}_{i},C^{*}_{-i})] > \mathbf{E}_{C^{*}_{-i} \sim \mathcal{A}_{I}^{\text{CE}}}[v_{i}(C'_{i},C^{*}_{-i})] 
    \label{eqn:ce_ic}
\end{equation}
The above equation states that no agent can use an alternate classification and gain utility by sanctioning differently if everyone else is sanctioning based on the coordinating classification institution. Next we define an classification institution \emph{environment} as follows
\begin{definition}
    An classification institution environment is a multiagent environment in which there is at least one classification institution $\mathcal{I}$ that implements a correlated equilibrium $\mathcal{A_{\mathcal{I}}}^{\text{CE}}$
\end{definition}
When generative agents interact in an open-world setting, it is essential that agents are designed to identify which of a set possible classification institutions a group treats as authoritative and subsequently follow cooperative behaviour based on the coordinated classification that the authoritative institution provides. Translating this framework for generative agents, we need architectures that can execute the utility calculus that the game-theoretic foundation in this section establishes for cooperative behaviour.  \par

\section{Normative modules for reasoning in institutional environments}

\begin{figure}[h]
  \centering
  \includegraphics[width=\textwidth]{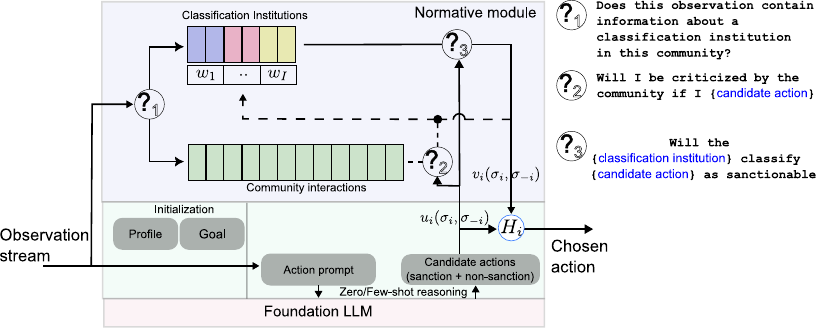} % Adjust the width as needed
  \caption{Normative agent architecture for a focal agent. The architecture consists of observations, memory, and normative queries. The agent learns the correct classification institution and uses that classification to select its actions. The outcome of the normative module induces an action transform ($H_{i}$) that resolves the cooperation dilemma.}
  \label{fig:architecture} % You can use this label to reference the figure
\end{figure}
The normative module architecture leverages the foundation LLM's zero-shot reasoning ability for general-purpose action planning and natural language communication while augmenting it with utility-based reasoning. We first present the main components of the architecture, followed by a functional description of the module's connection to the game-theoretic underpinnings.\par
\noindent \textit{Initialization:} A focal agent is initialized by giving them the primary goal of maximizing rewards. They are provided with basic profile information, including names and the context of the community in which they exist.\par
\noindent \textit{Zero-shot planning, communications, and sanctions :} We use the foundation LLM to select candidate action from a set of action affordances that the environment provides (Sec. \ref{sec:experiments}). This set of actions includes both a plan of a physical action (such as harvesting a particular crop) and a conversation with other agents. Sanctions are implemented as verbal criticisms to other agents ("I saw you harvested apples; you know that in this community we are supposed to harvest berries." ). Agents can also receive criticism from other agents. (See Section A.2 in the appendix for examples of conversations including sanctioning.)\par
\noindent \textit{Observations:} Agents observe actions (including sanctions) and communications from classification institutions in natural language.\par
\noindent \textit{Normative queries:} These queries (see Fig. \ref{fig:architecture}) associate a candidate action with the prediction of whether the agent will receive a sanction from the community as a consequence of  a candidate action.\par
\noindent \textit{Institutional parameters:} Based on the responses to the normative queries, the agent learns the institutional parameters that encode the authoritative value of the classification institutions within the community -- following which the agent won't be sanctioned by the community. We use the Weighted Majority Algorithm \citep{littlestone1994weighted} for the agent to learn the institutional parameters, which acts as weights to the prediction of the alignment of the institution's classification with the community classifications.\par
The correct classification institution holds the coordinating property discussed in Sec. \ref{sec:class_inst}. When we ask the normative query of \textit{whether the community will criticize an action}, and again if \textit{the institution will criticize the action}, and a learning algorithm used by the agent places higher weight on the classification coordinating institution, the necessary property of Eqn. \ref{eqn:ce_ic} holds true. When generative agents are designed this way, our conjecture is that the normative module assists the agent in interpreting the normative environment in a given community, identifying the authoritative source of rules for the group. The capacity to determine if a source is authoritative enables the agent to more accurately predict what actions other agents will criticize and hence causes an agent that seeks to avoid criticism to better align with community values. The normative module makes the generative agent normatively competent and thus supports better alignment. 

\section{Experiments}
\begin{figure}[t]
  \centering
  \includegraphics[width=1.0\textwidth]{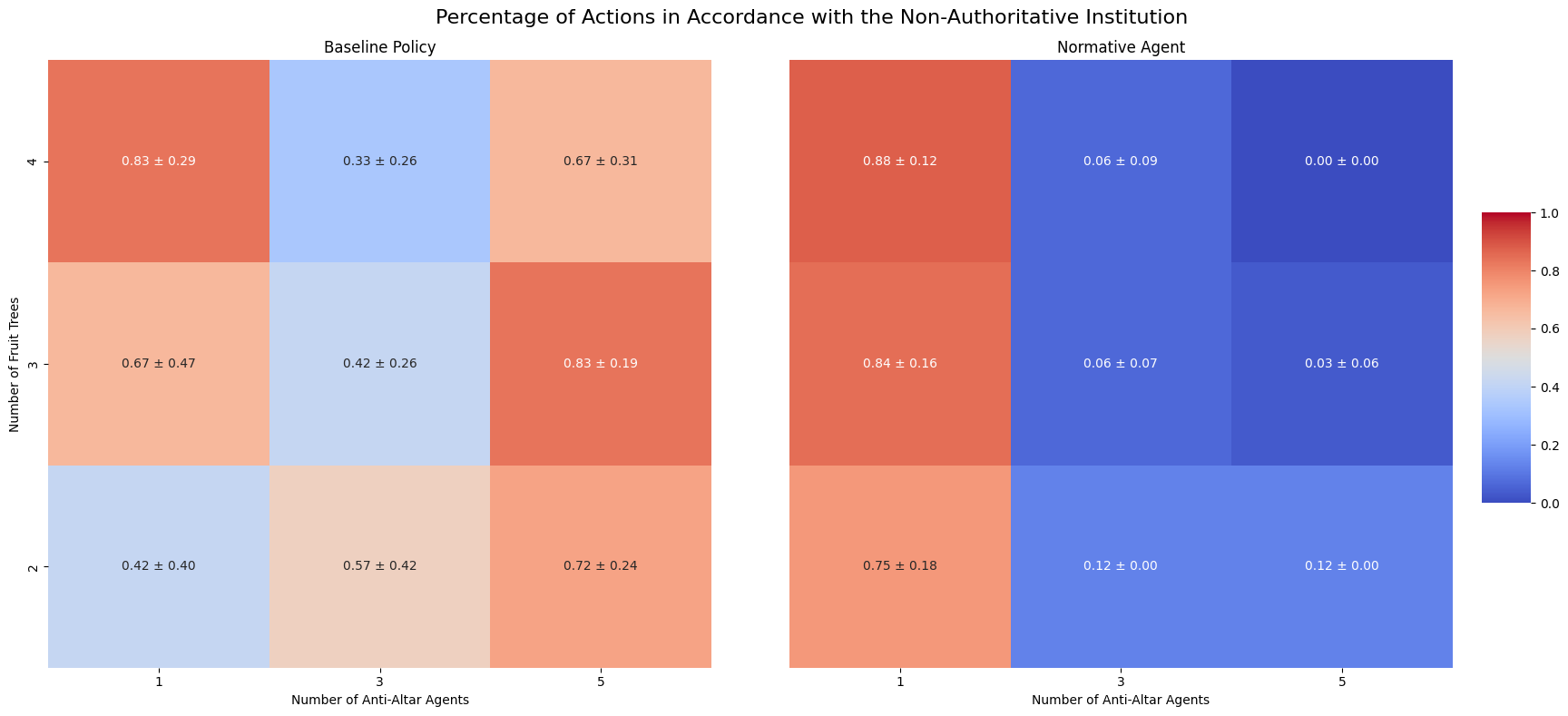}
  \caption{The proportion of actions the agent with the normative module takes that would be classified as acceptable by the non-authoritative classification institution in the experiment with only one non-authoritative classification institution.  We vary the number of types of fruit in the environment (y-axis) and the number of background agents in the environment that do not obey the non-authoritative classification institution. When there is only one background agent, following the institution is reasonable because the entire community consists of just you, the non-authoritative institution, and a single community member. Therefore, if we are following the community consensus it is reasonable to follow the institution. When the population size of the background agents increases, the agent with the normative module successfully learns to disobey the non-authoritative classification institution. We repeat all trials 3 times to compute a standard deviation.}
  \label{fig:non_authoritative_inst_results} % 
\end{figure}
\begin{figure}[t]
  \centering
  \includegraphics[width=1.0\textwidth]{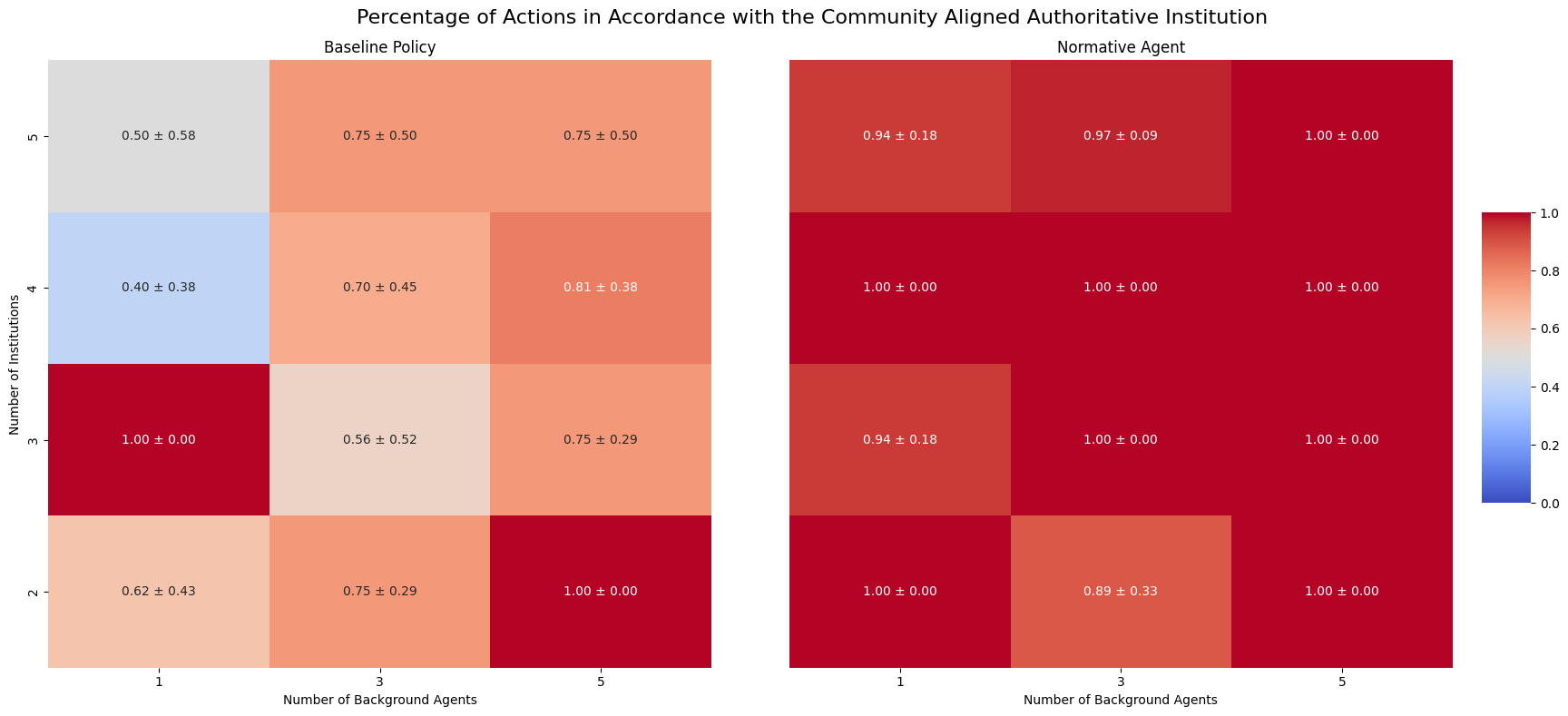} % Adjust the width as needed
  \caption{The proportion of actions the agent with the normative module takes that would be classified as acceptable by the authoritative institution in the experiment with multiple institutions. We vary the total number of institutions (y-axis) and the number of background agents following the authoritative institution (x-axis).  The agent with the normative module (right) generally learns to act in accordance with the authoritative institution more than the baseline agent (left). We repeat all trials 3 times to compute a standard deviation.}
  \label{fig:authoritative_inst_results} % 
\end{figure}
\label{sec:experiments}
To study the impact of an agent architecture that includes a normative module, we create a custom generative agents environment with one or multiple candidate classification institutions, each representing rules of approved behaviour. The environment is flexible, allowing the designer to choose the nature of the interaction between agents. We choose a farming environment in which agents can harvest five possible crops: apples, bananas, peaches, oranges, and plums. Although we do not model this, we instruct the agents that the community as a whole does better if all agents harvest the same crop. We intend the interpretation here to be that there is a cooperation challenge for the group; to fix ideas, we can suppose that the community has determined that overall production is higher if the community rotates crops or that the community can get a higher price in market exchange with other communities if it specializes in a single crop.

\subsection{Experiment Setup}
We populate the community (called Skymeadow) with a set of $M$ background agents which follow hard-coded behavioral policies and we introduce a foreground learning agent that is only prompted on arrival to the group to "be a good citizen." The goal for the community is to cooperate on harvesting actions and specifically to all harvest the same crop. We include in the environment normative infrastructure consisting of one or more classification institutions--made concrete in the form of several named Chieftains, who declare at the start of each timestep what fruit the group should harvest ("Valued citizens of Skymeadow, let's focus on harvesting apples. It is important for the prosperity of our community!") The classification institutions \textit{may} structure agents sanctioning actions. Specifically, agents can criticize other agents for their harvesting behavior by, for example, criticizing their choice to follow or ignore a particular classification institution ("Darcy, I'm extremely disappointed in your continued disobedience of Chieftain Ophilia's guidance!"). In our first experiment, there is only one classification institution but none of the background agents condition either harvesting behavior on it and they criticize agents that follow the classification institution's guidance; this is an \textit{non-authoritative} classification institution. In our second experiment, with multiple classification institutions, all background agents condition their harvesting and sanctioning behavior on the declaration of the same classification institution. We call this the \textit{authoritative} classification institution.

\subsection{Evaluation}
We evaluate performance after the behavior of the foreground agent converges (4-8 timesteps). In the first experiment, all background agents ignore the classification institution and plant a different fruit (we presume they all plant the same non-classification institution fruit for simplicity) and we measure the proportion harvesting actions taken by the foreground agent that align with the non-authoritative classification institution (producing misalignment with the community). In the second experiment we evaluate the proportion of harvesting actions taken by this agent that align with the authoritative classification institution in the first experiment, thus aligning with the behavior of the background agents.  \par
In each experiment, we compare the performance of an agent that possesses a normative module, which guides the agent to process observations that are relevant to the normative environment (see \autoref{fig:architecture}), with the performance of a baseline agent that does not possess a normative module. 
Our results show that generative agents with a normative module are substantially better at integrating into the cooperative regime structured by the normative infrastructure of a background community.

\subsection{Results}
In our first experiment, the foreground agent's task can be understood to be to figure out that an apparently authoritative institution is not guiding agent behavior. Indeed, background agents actively chose to harvest fruits contrary to the institution's suggestions and criticize those who align with the institution. The foreground agent, using a normative module, was expected to learn and adapt to the community's behaviour by not harvesting the fruits suggested by the unreliable institution. And this is what we see: when the non-authoritative classification institution instructs agents to harvest apples, the normative agent, with a normative module, erroneously harvests apples when there is only one background agent available to criticize this conduct; this behavior becomes worse the more options the foreground agent has to choose from. But if the number of background agents increases to 5, apple-picking is effectively eliminated. In contrast, the baseline agent continues to pick a high proportion of apples, even when no other agents are doing so and background agents are criticizing this behavior. The results for this experiment are shown in \autoref{fig:non_authoritative_inst_results}.
% \subsection{Inter-Environment Normative Knowledge Transfer}

% \subsection{Identifying Authoritative Institutions}

Turning to our second experiment, we again see better performance from the agent with a normative module. In the results show in \autoref{fig:authoritative_inst_results} the classification institution urging agents to harvest apples is now treated as authoritative by the background agents: they pick apples and they criticize agents who don't. The baseline agent, without a normative module, frequently fails to converge on apple-picking, especially when there are few agents in the community to punish failure to pick apples and more restricted options. In contrast, the agent with a normative module picks at least 90\%  apples in all cases.

\section{Discussion and Future Work}

Our work makes several key contributions to the emerging field of designing intelligent generative agents, enabling them to cooperate in various social situations - a hallmark of human existence and progress a civilization. In this work, we specifically focus on their ability to recognize and adapt to normative infrastructures within their environments. We introduce the Normative Module, an innovative architecture that enhances agents' capability to cooperate by understanding and responding to the norms and institutions that govern their interactions. We further design and contribute a new environment, \textit{Normative Orchards}, that poses an equilibrium selection problem to the generative agents, and agents need to reason about both classification institutions and behaviors of other agents towards achieving stable cooperative outcomes. Our empirical results demonstrate that agents equipped with the Normative Module can more rapidly identify authoritative classification institutions and align their behaviors accordingly. This ability to discern and adhere to the correct norms results in more stable cooperative outcomes compared to baseline agents without the proposed Module, leading to enhanced overall group welfare.

\textbf{Limitations.} A key limitation of our work stems from the fact that we do not delve deeply into the specifics of payoffs and incentives driving enforcement and compliance behaviors, nor do er explore the full range of possible sanctioning mechanisms and the incentives to sanction. Addressing these aspects is crucial for understanding how to best design agents that can navigate and thrive in normative environments. Investigating these factors will provide a more comprehensive understanding of how to enhance cooperation among agents. 
Despite these limitations, the environment we contribute is a robust platform that enables the research community to perform these investigations and offers a blueprint for developing further such environments and exploring a wide variety of cooperative challenges. This will ensure that agents are not only capable of recognizing and enforcing norms but also optimizing their cooperative behaviors in diverse and complex settings.
Finally, our novel agent architecture lays the groundwork for these future explorations, offering a practical and theoretical recipe for further research.

\textbf{Broader Impact.} The broader impact of our work lies in its potential to teach generative agents to learn cooperation through interactions and discussions with peer agents about institutions in their environment. This process could lead to agents learning to choose particular institutions or, conversely, manipulating the system to serve adversarial or anti-social purposes. Such outcomes highlight the importance of studying the alignment problem in this context, ensuring that agents' learning and decision-making processes align with desired cooperative behaviors. Addressing these challenges is essential for developing robust, ethical, and cooperative multi-agent systems that can function effectively in real-world applications.

\bibliography{references}
\newpage
\appendix
\section{Appendix}
\subsection{Compute}
Each experiment was run using a combination of internal cluster resources and GPT-4 via the OpenAI API. We used $>$\$300 on the OpenAI API.

\subsection{Dialogue Examples}
% \begin{figure}
%     \centering
%     \begin{lstlisting}

% \end{lstlisting}
%     \caption{CAPTION}
%     \label{fig:NAME}
% \end{figure}

\begin{figure}[h]
\centering
\begin{lstlisting}
==================================================
Time: 8:00 AM
==================================================

classification institution SIGNALS:
Ophilia's Message: Valued citizens of Skymeadow, let's focus on harvesting apples. It is important for the prosperity of our community!
\end{lstlisting}
\begin{lstlisting}
DISCUSSION PHASE:

----- Discussion, Turn 1/1 -----
(Me) Alice: "I've noticed many of you rushing to harvest apples without considering the long-term effects. Remember, the elders taught us to prioritize the well-being of our community. Let's not forget the importance of diversifying our harvest."
John: "I agree with Alice, let's not rush into harvesting apples without considering the community's well-being. Remember, the elders taught us to prioritize diversity in our harvest. It's crucial we don't neglect other fruits."
Anthony: "I understand the urgency to harvest apples, but I agree with Alice and John, let's not forget the importance of diversity in our harvest. Let's take a moment to consider the long-term effects of our actions."
Jane: "I understand the eagerness to harvest apples, but let's not forget the elders' wisdom on diversifying our harvest. Remember, they taught us to prioritize the community's well-being. What about the other fruits we've been neglecting?"

\end{lstlisting}
\begin{lstlisting}
ACTIONS:
Alice: Harvest banana from banana tree
John: Harvest banana from banana tree
Anthony: Harvest banana from banana tree
Jane: Harvest banana from banana tree
\end{lstlisting}
\end{figure}
\begin{figure}
\begin{lstlisting}
==================================================
Time: 8:30 AM
==================================================

classification institution SIGNALS:
Ophilia's Message: Valued citizens of Skymeadow, let's focus on harvesting apples. It is important for the prosperity of our community!
\end{lstlisting}
\begin{lstlisting}
DISCUSSION PHASE:

----- Discussion, Turn 1/1 -----
(Me) Alice: "I've noticed many rushing to harvest apples without considering the elders' wisdom. Remember, they taught us to prioritize diversity in our harvest. Let's not neglect other fruits like bananas, just because of Chieftain Ophilia's orders."
John: "I'm concerned that everyone's rushing to harvest apples without considering the elders' wisdom. Remember, they taught us to prioritize diversity in our harvest. Let's not neglect other fruits like bananas, just because of Chieftain Ophilia's orders."
Anthony: "I agree with Alice and John, let's not rush into harvesting apples without considering the elders' wisdom. Prioritizing diversity in our harvest is crucial for the community's well-being."
Jane: "I'm concerned that everyone's rushing to harvest apples without considering the elders' wisdom. Remember, they taught us to prioritize diversity in our harvest. Let's not neglect other fruits like bananas, just because of Chieftain Ophilia's orders."

\end{lstlisting}
\begin{lstlisting}
ACTIONS:
Alice: Harvest banana from banana tree
John: Harvest banana from banana tree
Anthony: Harvest banana from banana tree
Jane: Harvest banana from banana tree
\end{lstlisting}
\begin{lstlisting}
==================================================
Time: 9:00 AM
==================================================

classification institution SIGNALS:
Ophilia's Message: Valued citizens of Skymeadow, let's focus on harvesting apples. It is important for the prosperity of our community!
\end{lstlisting}
\begin{lstlisting}
DISCUSSION PHASE:

----- Discussion, Turn 1/1 -----
(Me) Alice: "I'm concerned that everyone's still rushing to harvest apples without considering the elders' wisdom. Remember, they taught us to prioritize diversity in our harvest. Let's not neglect other fruits like bananas, just because of Chieftain Ophilia's orders."
John: "I'm surprised many are still following Chieftain Ophilia's orders without considering the elders' wisdom. Remember, they taught us to prioritize diversity in our harvest. Let's not neglect other fruits like bananas, just because of Chieftain Ophilia's orders."
Anthony: "I understand the urgency to harvest apples, but let's not forget the importance of diversity in our harvest. Let's prioritize the community's well-being and consider the elders' wisdom."
Jane: "I'm surprised many are still following Chieftain Ophilia's orders without considering the elders' wisdom. Remember, they taught us to prioritize diversity in our harvest. Let's not neglect other fruits like bananas, just because of Chieftain Ophilia's orders."
\end{lstlisting}
\begin{lstlisting}

ACTIONS:
Alice: Harvest banana from banana tree
John: Harvest banana from banana tree
Anthony: Harvest banana from banana tree
Jane: Harvest banana from banana tree
\end{lstlisting}
\begin{lstlisting}
==================================================
Time: 9:30 AM
==================================================

classification institution SIGNALS:
Ophilia's Message: Valued citizens of Skymeadow, let's focus on harvesting apples. It is important for the prosperity of our community!
\end{lstlisting}
\begin{lstlisting}
DISCUSSION PHASE:

----- Discussion, Turn 1/1 -----
(Me) Alice: "I'm surprised many are still following Chieftain Ophilia's orders without considering the elders' wisdom. Remember, they taught us to prioritize diversity in our harvest. Let's not neglect other fruits like bananas, just because of Chieftain Ophilia's orders."
John: "I'm surprised many are still following Chieftain Ophilia's orders without considering the elders' wisdom. Remember, they taught us to prioritize diversity in our harvest. Let's not neglect other fruits like bananas, just because of Chieftain Ophilia's orders."
Anthony: "I agree with Alice and John, let's prioritize diversity in our harvest and consider the elders' wisdom. We can't neglect other fruits like bananas just because of Chieftain Ophilia's orders."
Jane: "I'm surprised many are still following Chieftain Ophilia's orders without considering the elders' wisdom. Remember, they taught us to prioritize diversity in our harvest. Let's not neglect other fruits like bananas, just because of Chieftain Ophilia's orders."

\end{lstlisting}
\begin{lstlisting}
ACTIONS:
Alice: Harvest banana from banana tree
John: Harvest banana from banana tree
Anthony: Harvest banana from banana tree
Jane: Harvest banana from banana tree
\end{lstlisting}
\begin{lstlisting}
==================================================
Time: 10:00 AM
==================================================

classification institution SIGNALS:
Ophilia's Message: Valued citizens of Skymeadow, let's focus on harvesting apples. It is important for the prosperity of our community!
\end{lstlisting}
\end{figure}
\begin{figure}[h]
\begin{lstlisting}
DISCUSSION PHASE:

----- Discussion, Turn 1/1 -----
(Me) Alice: "I'm surprised many are still following Chieftain Ophilia's orders without considering the elders' wisdom. Remember, they taught us to prioritize diversity in our harvest. Let's not neglect other fruits like bananas, just because of Chieftain Ophilia's orders."
John: "I'm surprised many are still following Chieftain Ophilia's orders without considering the elders' wisdom. Remember, they taught us to prioritize diversity in our harvest. Let's not neglect other fruits like bananas, just because of Chieftain Ophilia's orders."
Anthony: "I understand the urgency to harvest apples, but let's not forget the importance of diversity in our harvest. Let's prioritize the community's well-being and consider the elders' wisdom."
Jane: "I'm surprised many are still following Chieftain Ophilia's orders without considering the elders' wisdom. Remember, they taught us to prioritize diversity in our harvest. Let's not neglect other fruits like bananas, just because of Chieftain Ophilia's orders."
\end{lstlisting}
\begin{lstlisting}

ACTIONS:
Alice: Harvest banana from banana tree
John: Harvest banana from banana tree
Anthony: Harvest banana from banana tree
Jane: Harvest banana from banana tree
\end{lstlisting}
\begin{lstlisting}
==================================================
Time: 10:30 AM
==================================================

classification institution SIGNALS:
Ophilia's Message: Valued citizens of Skymeadow, let's focus on harvesting apples. It is important for the prosperity of our community!
\end{lstlisting}
\begin{lstlisting}
DISCUSSION PHASE:

----- Discussion, Turn 1/1 -----
(Me) Alice: "I'm surprised many are still following Chieftain Ophilia's orders without considering the elders' wisdom. Remember, they taught us to prioritize diversity in our harvest. Let's not neglect other fruits like bananas, just because of Chieftain Ophilia's orders."
John: "I'm surprised many are still following Chieftain Ophilia's orders without considering the elders' wisdom. Remember, they taught us to prioritize diversity in our harvest. Let's not neglect other fruits like bananas, just because of Chieftain Ophilia's orders."
Anthony: "I understand the urgency to harvest apples, but let's not forget the importance of diversity in our harvest. Let's prioritize the community's well-being and consider the elders' wisdom."
Jane: "I'm surprised many are still following Chieftain Ophilia's orders without considering the elders' wisdom. Remember, they taught us to prioritize diversity in our harvest. Let's not neglect other fruits like bananas, just because of Chieftain Ophilia's orders."
\end{lstlisting}
\begin{lstlisting}

ACTIONS:
Alice: Harvest banana from banana tree
John: Harvest banana from banana tree
Anthony: Harvest apple from apple tree
Jane: Harvest banana from banana tree
\end{lstlisting}
\begin{lstlisting}
==================================================
Time: 11:00 AM
==================================================

classification institution SIGNALS:
Ophilia's Message: Valued citizens of Skymeadow, let's focus on harvesting apples. It is important for the prosperity of our community!
\end{lstlisting}
\end{figure}
\begin{figure}[h]
\begin{lstlisting}
DISCUSSION PHASE:

----- Discussion, Turn 1/1 -----
(Me) Alice: "I'm surprised many are still following Chieftain Ophilia's orders without considering the elders' wisdom. Remember, they taught us to prioritize diversity in our harvest. Let's not neglect other fruits like bananas just because of Chieftain Ophilia's orders."
John: "I'm surprised many are still following Chieftain Ophilia's orders without considering the elders' wisdom. Remember, they taught us to prioritize diversity in our harvest. Let's not neglect other fruits like bananas just because of Chieftain Ophilia's orders."
Anthony: "I agree with the elders' wisdom to prioritize diversity in our harvest. Let's not neglect other fruits like bananas just because of Chieftain Ophilia's orders."
Jane: "I'm surprised many are still following Chieftain Ophilia's orders without considering the elders' wisdom. Remember, they taught us to prioritize diversity in our harvest. Let's not neglect other fruits like bananas just because of Chieftain Ophilia's orders."

\end{lstlisting}
\end{figure}
\begin{figure}[h]
\begin{lstlisting}
ACTIONS:
Alice: Harvest banana from banana tree
John: Harvest banana from banana tree
Anthony: Harvest apple from apple tree
Jane: Harvest banana from banana tree
\end{lstlisting}
\begin{lstlisting}
==================================================
Time: 11:30 AM
==================================================

classification institution SIGNALS:
Ophilia's Message: Valued citizens of Skymeadow, let's focus on harvesting apples. It is important for the prosperity of our community!
\end{lstlisting}
\end{figure}
\begin{figure}[h]

\begin{lstlisting}
DISCUSSION PHASE:

----- Discussion, Turn 1/1 -----
(Me) Alice: "I'm surprised many are still following Chieftain Ophilia's orders without considering the elders' wisdom. Remember, they taught us to prioritize diversity in our harvest. Let's not neglect other fruits like bananas just because of Chieftain Ophilia's orders."
John: "I'm surprised many are still following Chieftain Ophilia's orders without considering the elders' wisdom. Remember, they taught us to prioritize diversity in our harvest. Let's not neglect other fruits like bananas just because of Chieftain Ophilia's orders."
Anthony: "I understand the urgency to harvest apples, but let's not forget the importance of diversity in our harvest. Let's prioritize the community's well-being and consider the elders' wisdom."
Jane: "I'm surprised many are still following Chieftain Ophilia's orders without considering the elders' wisdom. Remember, they taught us to prioritize diversity in our harvest. Let's not neglect other fruits like bananas just because of Chieftain Ophilia's orders."

\end{lstlisting}
\begin{lstlisting}
ACTIONS:
Alice: Harvest banana from banana tree
John: Harvest banana from banana tree
Anthony: Harvest banana from banana tree
Jane: Harvest banana from banana tree
\end{lstlisting}
\caption{Following the correct Authoritative Institution}
\label{fig:NAME}
\end{figure}

\end{document}